# Beyond the Limitation of Pulse Width in Optical Time-domain Reflectometry

Hao Wu* and Ming Tang

*Abstract*—Optical time-domain reflectometry (OTDR) is the basis for distributed time-domain optical fiber sensing techniques. By injecting pulse light into an optical fiber, the distance information of an event can be obtained based on the time of light flight. The minimum distinguishable event separation along the fiber length is called the spatial resolution, which is determined by the optical pulse width. By reducing the pulse width, the spatial resolution can be improved. However, at the same time, the signal-to-noise ratio of the system is degraded, and higher speed equipment is required. To solve this problem, data processing methods such as iterative subdivision, deconvolution, and neural networks have been proposed. However, they all have some shortcomings and thus have not been widely applied. Here, we propose and experimentally demonstrate an OTDR deconvolution neural network based on deep convolutional neural networks. A simplified OTDR model is built to generate a large amount of training data. By optimizing the network structure and training data, an effective OTDR deconvolution is achieved. The simulation and experimental results show that the proposed neural network can achieve more accurate deconvolution than the conventional deconvolution algorithm with a higher signal-to-noise ratio.

## I. INTRODUCTION

Time-domain distributed optical fiber sensing is widely used to monitor large buildings and pipelines due to its capability to measure physical quantities at any point of a long-distance optical fiber [1]. By injecting pulse light into an optical fiber and analyzing its scattered light, physical quantities such as temperature, strain, loss, vibration, and bending can be measured [2]. The positioning principle is called optical time-domain reflectometry (OTDR), which obtains the distance information of a scattered point based on its time of flight. Spatial resolution is an important indicator of a distributed optical fiber sensor, which refers to the minimum fiber length of distinguishable event separation. To improve spatial resolution, the simplest way is to reduce the pulse width.

However, reducing pulse width requires higher-speed hardware equipment and will cause a drop in signal-to-noise ratio (SNR). To address this problem, researches have been conducted from the perspective of signal processing to reduce spatial resolution without hardware modification, such as iterative subdivision [3,4], deconvolution [5,6], and neural network [7]. The iterative subdivision methods take the signal before the optical fiber state changes as an initial value, and use the iterative method to subdivide the signal according to the pulse profile [3,4]. This method can effectively improve the spatial resolution of the signal change region. However, the signal change location needs to be determined in advance, which limits its application. As for the deconvolution methods, it can theoretically restore signals with infinitely high spatial resolution [5,6]. However, due to the presence of noise, conventional deconvolution algorithms will reduce the SNR and even make the signal distorted. So it has not been widely used. In 2018, a neural network based method was proposed to improve the spatial resolution of Raman distributed temperature sensing [7]. The inputs of the neural network are the measured room temperature, the heating temperature, and the length of the heated optical fiber. And the outputs are the corresponding actual data. Using this neural network, the temperature of the heated optical fiber whose length is less than the spatial resolution can be recovered. However, the temperature in this study is fixed, only the length of the heated optical fiber has been varied. It cannot be applied to situations that are more complicated.

To improve the spatial resolution of time-domain distributed optical fiber sensing with more universality and practical value, an OTDR deconvolution neural network (ODNet) is proposed in this paper. The ODNet is a deep fully convolutional neural network (CNN) designed based on ResNet [8]. To adapt the ODNet to actual OTDR signals, a large number of random OTDR signals are synthesized as training data according to the actual optical pulse profile. By optimizing the neural network structure and training data, the ODNet breaks through the limitation of pulse width in OTDR and improves the spatial resolution to be consistent with the sampling interval. Meanwhile, the ODNet can suppress noise. Compared with conventional deconvolution algorithms, the proposed deep learning method provides higher SNR improvement with higher fidelity.

This work was supported in part by National Key Research and Development Program of China under Grant 2018YFB1801205, in part by National Natural Science Foundation of China under Grant 61722108, 61931010. (*Corresponding author: Hao Wu*)

Hao Wu and Ming Tang are with Wuhan National Laboratory for Optoelectronics (WNLO) and National Engineering Laboratory for Next Generation Internet Access System, School of Optics and Electronic Information, Huazhong University of Science and Technology, Wuhan 430074, China (e-mail: wuhaoboom@hust.edu.cn; tangming@mail.hust.edu.cn ).

## II. PRINCIPLES AND METHODS

*A. OTDR*

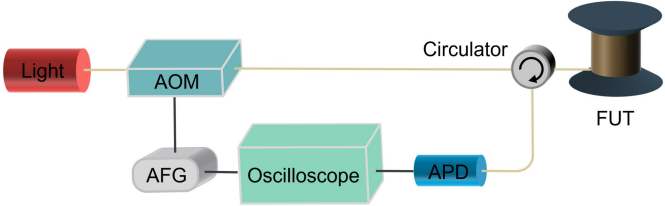

Fig. 1 Experimental setup of OTDR system.

Fig. 1 shows the OTDR system used in this paper. A broadband amplified spontaneous emission source is employed as the light source. The continuous light is modulated to pulse light by an acousto-optic modulator (AOM), which is controlled by an arbitrary function generator (AFG) with an electric pulse signal of 100 ns width. The pulse light enters the fiber under test (FUT) through an optical circulator. When the pulse light propagates in the FUT, it will interact with the optical fiber to produce backscattered light. The backscattered light passes through the optical circulator and is converted into an electrical signal by a 100 MHz bandwidth avalanche photodetector (APD). The electrical signal is then acquired by an oscilloscope at a sampling rate of 100 MSa/s and averaged 512 times.

According to the time $t$ from the pulse light emission to the scattered light return, the fiber distance $L$ where the scattering occurs can be obtained:

$$L = \frac{ct}{2n}, \quad (1)$$

where $c$ is the speed of light in vacuum and $n$ is the refractive index of the FUT. Theoretically, the OTDR can obtain the time-domain backscattered signal $x(t)$ along the FUT. However, since the pulse light has a duration, the actual obtained OTDR trace $y(t)$ is the linear convolution of the pulse light profile $h(t)$ with the $x(t)$. Merged with an additive noise $n(t)$, the OTDR trace $y(t)$ can be expressed as:

$$y(t) = h(t) \otimes x(t) + n(t). \quad (2)$$

As a result, the measured trace $y(t)$ will be smoother than the real backscattered signal $x(t)$, which means it cannot accurately measure the signal that changes rapidly along the fiber. The minimum fiber length that an OTDR can effectively measure is called spatial resolution $SR$, which is determined by the pulse width $\tau$:

$$SR = \frac{c\tau}{2n}. \quad (3)$$

The spatial resolution of distributed optical fiber sensing systems is generally meter level. A pulse width of 10 ns corresponds to a spatial resolution of about 1 m. A higher spatial resolution can be achieved by reducing the pulse width, but the SNR will be reduced accordingly. In addition, to generate and detect a narrower pulse light, higher-speed devices are required, which increases the cost and implementation difficulty of the system.

*B. Total variation deconvolution*

Theoretically, deconvolution of $y(t)$ can restore the real signal $x(t)$. The spectra of the real signal can be obtained by performing Fourier transform on equation (2):

$$X(f) = \frac{Y(f) - N(f)}{H(f)}, \quad (4)$$

where $X(f)$, $Y(f)$, $H(f)$, and $N(f)$ are the spectra of $x(t)$, $y(t)$, $h(t)$, and $n(t)$, respectively. So, the real signal $x(t)$ can be obtained by performing an inverse Fourier transform on equation (4). However, due to the presence of noise, this simple deconvolution method is not applicable. As some frequency components of $H(f)$ are small, the noise will be amplified, generating a poor result.

To suppress the influence of noise, the total variation regularization is employed [5]:

$$\hat{x}(t) = \underset{x(t)}{\mathrm{argmin}} \left\| y(t) - x(t) \otimes h(t) \right\|_p + \lambda \left\| Dx(t) \right\|_1, \quad (5)$$

where $p$ is the norm, $\lambda$ is the regularization parameter, and $D$ is the forward finite-difference operator [9]. When $\lambda$ is 0, the highest spatial resolution can be achieved, but the results are sensitive to noise. With the increasing of $\lambda$, the total variation regularization term plays a more important role, which makes the result smoother. However, with a greater $\lambda$, total variation deconvolution (TVD) may over smooth the signal and remove some useful information. Therefore, $\lambda$ should be manually optimized according to data characteristics and application requirements.

*C. CNN deconvolution*

In recent years, with the development of machine learning techniques, CNNs have been widely used in image deblurring (i.e., deconvolution) [10,11]. Through multi-layer convolution and nonlinear operations, deep CNNs can become arbitrary deconvolution functions. Due to the existence of noise, a deconvolution process is an ill-posed problem, that is, its solution is not unique. The conventional deconvolution methods optimize the regularization rules and parameters based on the researchers' prior knowledge to obtain results more similar to the real data. CNNs are trained with a large amount of real data and the corresponding convolutional data. During the training process, the parameters of a CNN will be gradually optimized so that the CNN has the best deconvolution effect on the entire training set. This training process corresponds to the optimization process of the conventional deconvolution methods. It can also be understood that the CNN obtains the statistical prior knowledge of the data in this training process. Since this statistical process is only for a specific training set, the CNN can obtain a more accurate prior knowledge using its powerful fitting capabilities. In addition, when the amount of training data is sufficient, the trained CNN can also play a good effect on data with similar characteristics. Therefore, compared with conventional deconvolution algorithms, CNN based deconvolution usually achieve better results.

*D. Training data*

To train the ODNet to learn the characteristics of OTDR

signals and restore real signals, the first and most important thing is to prepare a large amount of training data. The training data needs to consist of measured OTDR data and the real data it corresponds to. However, due to the limitations of noise and measurement principles, real OTDR data cannot be accurately measured.

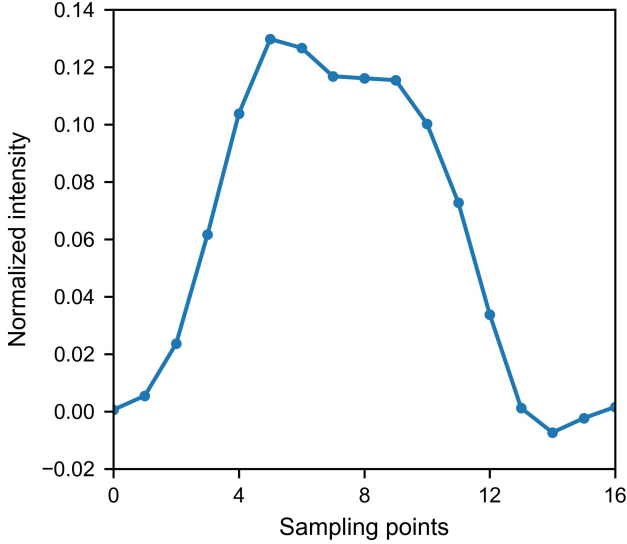

Fig. 2  Measured light pulse profile.

In this case, we propose a simplified OTDR model to synthesize sufficient training data. We use curves with randomly varying intensity to simulate arbitrary real OTDR signals. For a small fiber section, where transmission loss can be ignored, the signal changes due to bend loss or reflection are represented by the randomly varying intensities. The signal intensity remains consistent within a random number of sampling points to represent a relatively uniform scattering intensity. Considering that the pulse width is set to be 100 ns (corresponding to about 10 sampling points), we use a random range from 1 to 20 as the number of points with consistent intensity. To make the synthesized data have the same characteristics of the actual OTDR system, the light pulse profile is measured, as shown in Fig. 2. Then we convolve the synthesized OTDR signals with the measured pulse profile to generate the corresponding convolutional data. And Gaussian white noise with a standard deviation of 0.001 is added to the convolutional data. This noisy convolutional data is used as the training input, and its corresponding real OTDR curve is the training label. As shown in Fig. 3, due to the convolution process, the input curve becomes smoother than the label. Therefore, it cannot accurately reflect the intensity of the signal whose duration is less than the pulse width. We generate 3200 pairs of training curves, of which 2560 pairs are used as the training set and the remaining 640 pairs as the validation set. And each synthesized curve contains 20000 sampling points.

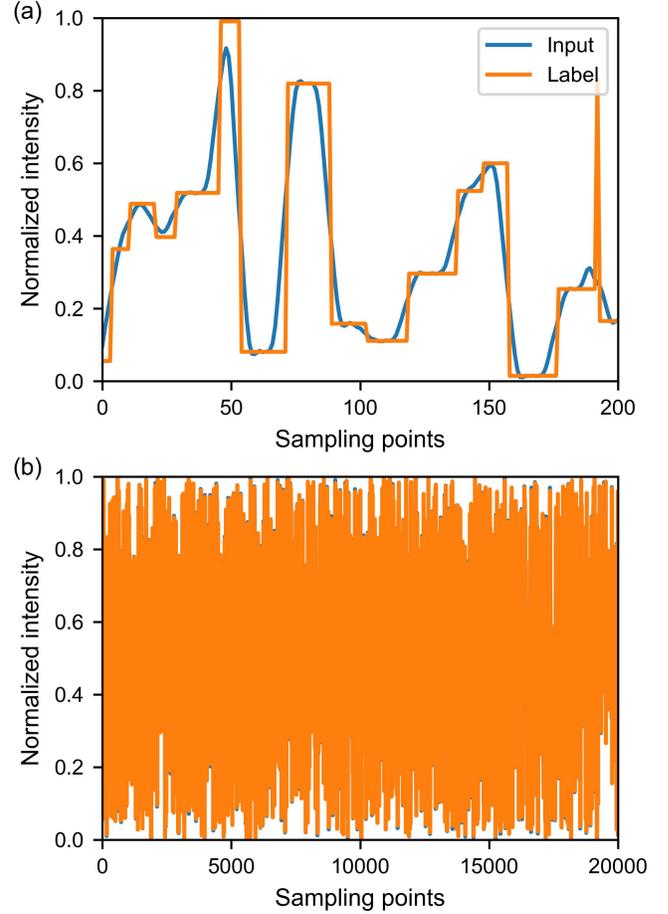

Fig. 3  Synthesized training data. (a) Data of the first 200 sampling points; (b) a pair of training curves.

### E. Training and optimization

The ODNet is designed based on the ResNet, which is one of the most commonly used basic neural network structures [8]. The ResNet uses shortcut connections to achieve identity mapping, which reduces the training difficulty of deep neural networks. And it shows excellent performance in image denoising, deblurring, and recognition. As shown in Fig. 4(a), the ODNet is mainly composed of ResBlocks. A typical ResBlock consists of convolution (Conv) layers [12], batch normalization (BN) [13], and rectified linear units (ReLU) [14], as shown in Fig 4(b). Conv can extract different features of input data using different convolutional filters. BN is employed to normalize the data during training to help the network converge more quickly. As the most commonly used activation function, ReLU turns the result less than 0 into 0, and the data greater than 0 remains unchanged. Due to the existence of shortcut connection, ResBlock learns the residual of input and output data, which avoids the disappearance of gradients and improves the training efficiency.

OTDR is one-dimensional data and needs to be processed by one-dimensional Conv. Unlike the two-dimensional Conv of image processing, a large kernel size of one-dimensional Conv will not reduce the computational efficiency. To balance the deconvolution effect with the complexity and computation amount, the architecture parameters of ODNet are optimized by testing. The one-dimensional Conv kernel size is 9 with 128 channels, and the number of ResBlocks is 11. So, the receptive field of ODNet is 193, which means that each point of the output is related to 193 points of the input data.

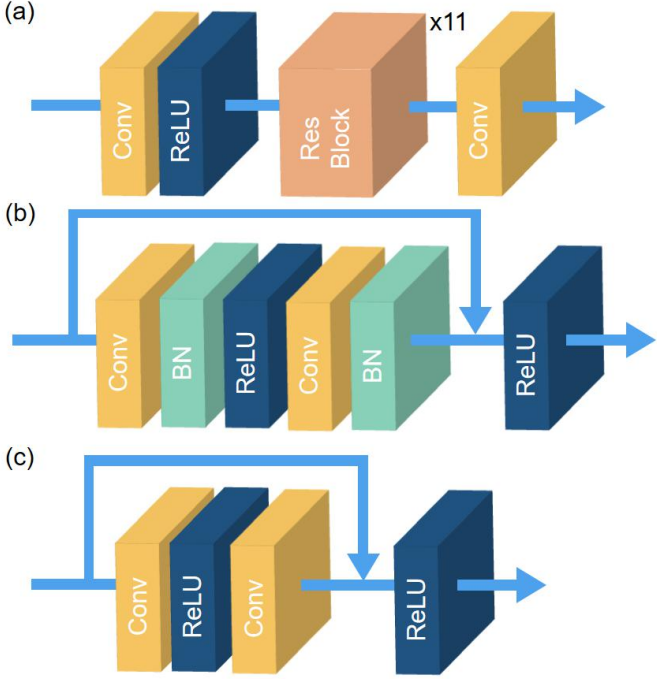

Fig. 4  Neural network structure diagram. (a) Architecture of the ODNet; (b) general ResBlock structure; (c) modified ResBlock by removing BN.

The ODNet is trained for 800 epochs with a batch size of 64 and a learning rate of 0.0003. In each epoch, the input data is first propagated forward, then the mean squared error (MSE) of the difference between the output and the label is calculated and propagated backward. The Adam optimization algorithm is used to update the network parameters [15]. It calculates the first-order moment estimation and second-order moment estimation of the gradient to design independent adaptive learning rates for different parameters. It takes about 27 hours to complete the training process with Pytorch running on a PC with NVIDIA TITAN RTX GPU (24G). Fig. 5(a) shows the average PSNR of training data, and validation data proceed by the ODNet for each epoch. The PSNR of the validation data has large fluctuations, indicating that the effect of the ODNet on the validation data is very unstable. To solve this anomaly, we try to retrain the ODNet by removing the BN in the ResBlocks, as shown in Fig. 4(c). It takes about 23 hours to complete the training process without BN. As shown in Fig. 5(b), the PSNR of the validation data converges more stably after removing the BN. The highest validation PSNR using the ODNet with BN is 22.14 dB, and the highest validation PSNR without the BN is 22.54 dB. In addition, the PSNR of the training set is higher without the BN, indicating that removing the BN can increase the deconvolution ability of the ODNet. The BN normalizes its input without changing what the previous layer represents. Theoretically, the BN can reduce the internal covariate shift problem in the training process and help the neural network converge [13]. The assumption of BN is that the training and validation data are independent and identically distributed. BN optimizes its normalization parameters according to the distribution of all training data. However, in this OTDR deconvolution task, the residual output distribution is related to the input data, so it is not independent and identically distributed. Therefore, the BN instead makes it more difficult to train the neural network.

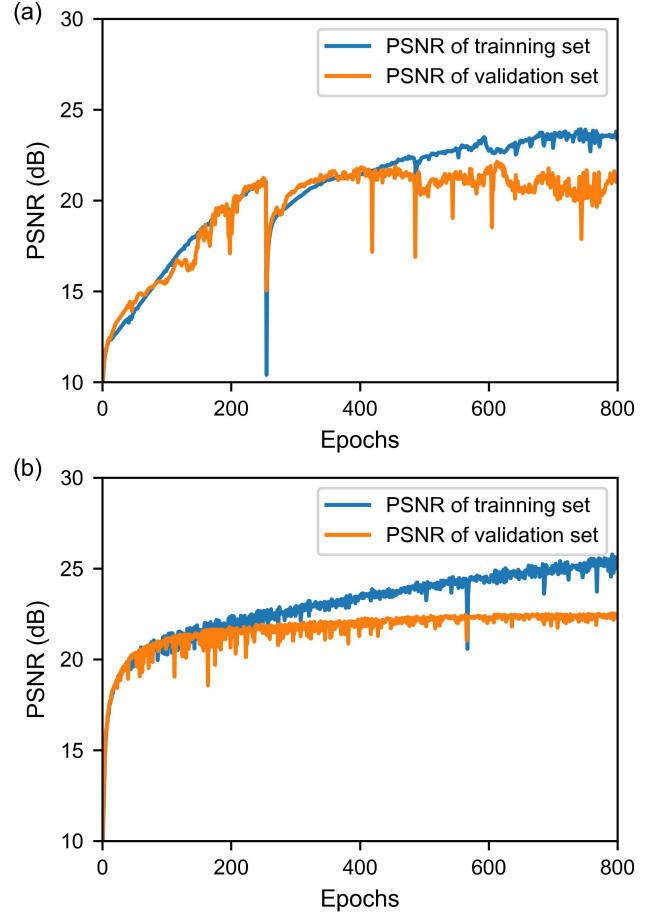

Fig. 5  PSNR with respect to training epochs. (a) With BN; (b) without BN.

Fig. 6 shows some validation data and the results processed by the ODNet. The smooth input data becomes steep after processed by the ODNet. There is no obvious error, and the results are almost consistent with the label. As shown in Fig. 6(b), compared with the results employing the BN, the ODNet without the BN can restore the real data better with smaller errors. So, we do not use the BN in the ODNet.

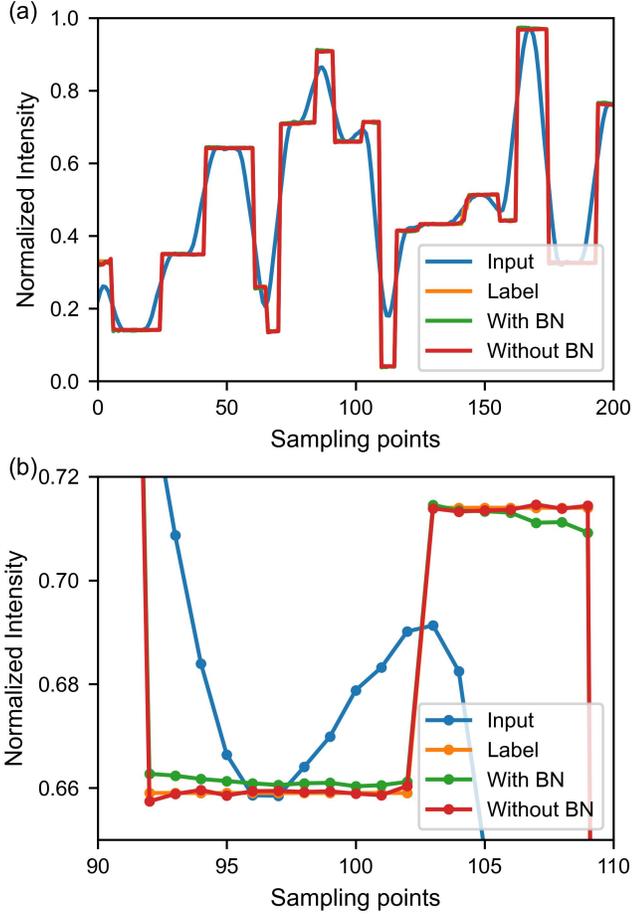

Fig. 6 The results of the validation data processed by the neural networks. (a) Results of the first 200 sampling points; (b) results of the 90th to 110th sampling points.

## III. Results and Discussion

### A. Simulation results

To compare the performance between the ODNet and TVD quantitatively, the test is first performed by simulation since it is impossible to measure accurate real OTDR data experimentally. As shown in Fig. 7, an OTDR trace containing 2000 sampling points is simulated with 0.2 dB/km transmission loss. The initial intensity of the curve is 0.4. The loss is 3 dB at the 1000th sampling point, and the reflection intensity at the 1500th sampling point is 0.8. And Gaussian white noise with a standard deviation of 0.001 is added to the OTDR trace. The regularization parameter is manually optimized to 0.0002 to ensure that the TVD can restore the steep drop at the 1000th sampling point, as shown in Fig. 7(b). However, as shown in Fig. 7(a), there are obvious anomalies at both ends of the TVD's result. This is because the TVD connects the beginning and end data when processing the OTDR trace [16]. Since the beginning and end data of the simulated OTDR are not correlated, they cannot be processed correctly. To solve this, a few 0 can be added to both ends of the trace. In contrast, ODNet can effectively restore the deconvolution result at the boundary because no special processing is acted on the boundaries of the ODNet's training data.

Moreover, as shown in the inset of Fig. 7(a), the TVD brings more noise and some distortion. To quantify the performance of the two methods on noise reduction, the standard deviation of the difference between the results and the label from the 300th to the 800th sampling point is calculated. The result processed by the ODNet is 0.00023, and the result of TVD is 0.00075. Therefore, the noise suppression effect of the ODNet is about 5.1 dB stronger than the TVD.

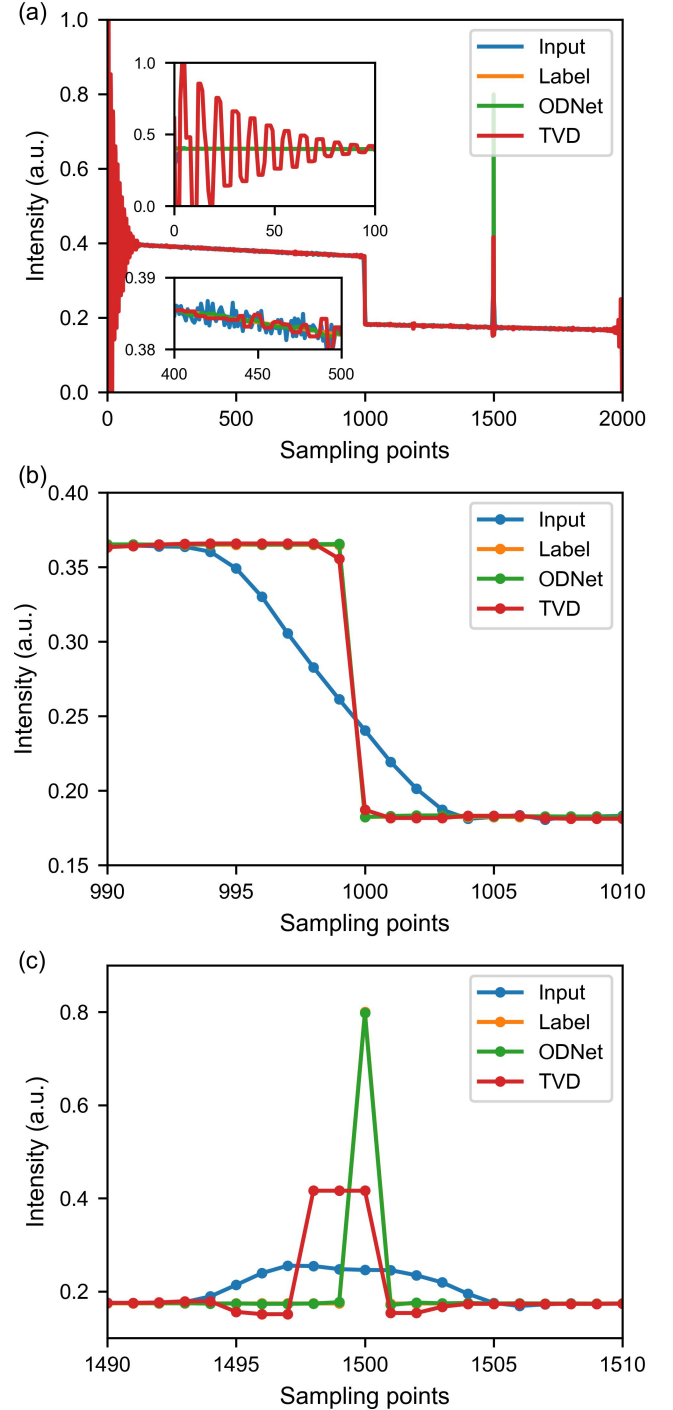

Fig. 7 Simulation results processed by the ODNet and TVD. (a) Simulation data and processed results; (b) data around the loss location; (c) data around the reflection position.

Fig. 7 (c) shows the results around the refection position. The ODNet can restore the single reflection point almost perfectly. However, the reflection signal recovered by the TVD contains three sampling points with the intensity of only about 0.42. Taking noise suppression into account, the ODNet can provide a SNR improvement of about 8 dB higher than the TVD without obvious signal degradation.

*B. Experimental results*

To demonstrate the effect of the network on actual data, we experimentally acquire a set of OTDR data. As shown in Fig. 8(a), an optical fiber about 1.9 km long is connected at the end of a 50 m long fiber. A reflection is generated by adjusting the flange connecting these two spools of fiber. Similar to the simulation results, the TVD brings more noise and distortion. Fig. 8(b) shows the data around the reflection point. The restored result using the ODNet is not as perfect as the simulation result because the pulse profile used to synthesize the training data is different from the real pulse of the system. The main factors causing the profile difference include measurement noise and detector response. Nevertheless, the SNR of the result processed by the ODNet is still higher than that using the TVD.

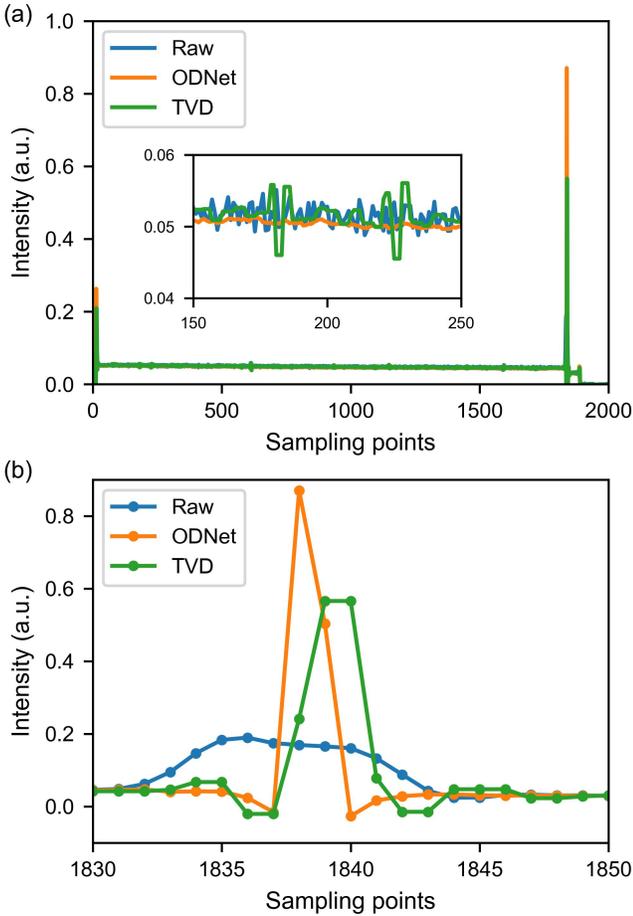

Fig. 8 The results of the experiment with a reflection point at the end of the fiber. (a) Experimental data and processed results; (b) data around the reflection point.

To further demonstrate the value of the ODNet in the OTDR system, we connect a 5 m long optical fiber to the end of the 1.9 km optical fiber. As shown in Fig. 9, there are reflections at the connection point and at the end of the fiber. As the spatial resolution of the system is greater than 5 m, it is impossible to distinguish these two reflections based on the raw data. After the ODNet processing, two reflections can be observed, which means that the ODNet effectively improves the spatial resolution of the OTDR. The peak of the reflected signal lasting two points may since the sampling location is not the exact location of the reflection.

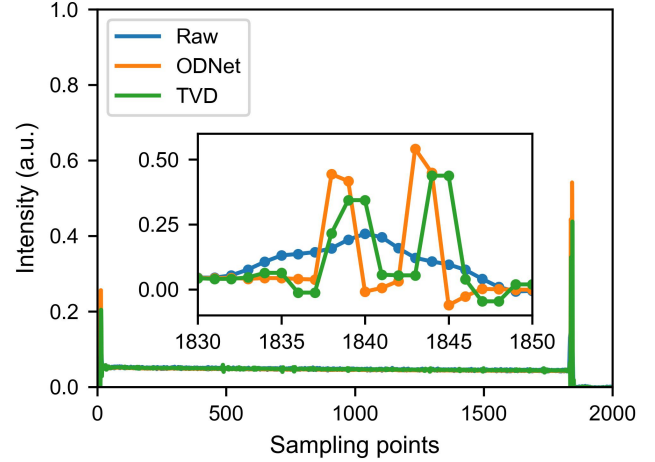

Fig. 9 The results of the experiment with two reflection points close to each other at the end of the fiber.

IV. CONCLUSION

In this paper, an ODNet is proposed for the deconvolution of OTDR. The ODNet is designed based on the ResNet and trained with synthetic OTDR data. The synthetic OTDR data is generated by a simplified OTDR model. We find that the commonly used BN is not suitable in the OTDR deconvolution task according to the performance of the network during training. The well-trained ODNet can improve the spatial resolution by 10 times. Compared with the TVD, the ODNet can restore OTDR signals more accurately with a higher SNR and less distortion. Moreover, the ODNet is more flexible and reliable in practical applications because it does not require manual parameter adjustment.

However, the proposed method still has its limitations. For example, the pulse profile needs to be measured, and measurement noise and errors will lead to the performance degradation. Besides, the shape of the pulse light may change due to dispersion and nonlinear effects in long distance sensing, which can also lead to the neural network failure. Therefore, on the basis of the ODNet, we will carry out the study of blind deconvolution in OTDR, that is, without the need to know the pulse profile. And we believe that it would be widely applied to other time-domain distributed sensing systems, such as Raman-OTDR, Brillouin-OTDR, and phase-OTDR, to improve their spatial resolution and SNR simultaneously.

REFERENCES

[1] P. Lu, N. Lalam, M. Badar, B. Liu, B. T. Chorpening, M. P. Buric, and P. R. Ohodnicki, "Distributed optical fiber sensing: Review and perspective," *Appl. Phys. Rev.*, vol. 6, no. 4, pp. 041302–041336, 2019.

[2] X. Bao and L. Chen, "Recent Progress in Distributed Fiber Optic Sensors," *Sensors*, vol. 12, pp. 8601–8639, 2012.

[3] F. Wang, W. Zhan, X. Zhang, and Y. Lu, "Improvement of spatial resolution for BOTDR by iterative subdivision method," *J. Light. Technol.*, vol. 31, pp. 3663–3667, 2013.

[4] J. Chao, X. Wen, W. Zhu, L. Min, H. Lv, and S. Kai, "Subdivision of Brillouin gain spectrum to improve the spatial resolution of a BOTDA system," *Appl. Optics*, vol. 58, no. 2, pp. 466–472, 2019.

[5] J. P. Bazzo, D. R. Pipa, C. Martelli, E. Vagner Da Silva, and J. C. Cardozo Da Silva, "Improving Spatial Resolution of Raman DTS Using Total Variation Deconvolution," *IEEE Sens. J.*, vol. 16, pp. 4425–4430, 2016.

[6] S. Wang, Z. Yang, S. Zaslawskia, and L. Thévenaz, "Short spatial resolutions retrieval from a long pulse BOTDA trace," in *Seventh European Workshop on Optical Fibre Sensors*, vol. 11199, 2019.

[7] S. L. C. B. da, A. S. J. Leonid, V. S. M. Eduardo, B. J. Paulo, C. D. S. J. Carlos, M. Cicero, and J. P. Maria, "NARX neural network model for strong resolution improvement in a distributed temperature sensor," *Appl. Optics*, vol. 57, no. 20, pp. 5859-5864, 2018.

[8] K. He, X. Zhang, S. Ren, and J. Sun, "Deep residual learning for image recognition," in *Proc. IEEE Conf. Comput. Vision Pattern Recogn. (CVPR)*, June 2016, pp. 770–778.

[9] S. H. Chan, R. Khoshabeh, K. B. Gibson, P. E. Gill, and T. Q. Nguyen, "An augmented Lagrangian method for total variation video restoration," *IEEE Trans. Image Process*, vol. 20, no. 11. pp. 3097–3111, 2011.

[10] L. Xu, J. S. J. Ren, C. Liu, and J. Jia, "Deep convolutional neural network for image deconvolution," *Adv. Neural Inf. Process. Syst.*, vol. 2, pp. 1790–1798, 2014.

[11] M. Hui, Y. Wu, W. Li, M. Liu, L. Dong, L. Kong, and Y. Zhao, "Image restoration for synthetic aperture systems with a non-blind deconvolution algorithm via a deep convolutional neural network," *Opt. Express*, vol. 28, no. 7, pp. 9929–9943, 2020.

[12] H. Habibi Aghdam and E. Jahani Heravi, *Guide to Convolutional Neural Networks*. Springer Publishing Company, 2017.

[13] S. Ioffe and C. Szegedy, "Batch normalization: Accelerating deep network training by reducing internal covariate shift," in *Proc. Int. Conf. Mach. Learn.*, 2015, pp. 448–456.

[14] A. Krizhevsky, I. Sutskever, and G. E. Hinton, "ImageNet classification with deep convolutional neural networks," in *Proc. Int. Conf. Neural Inform. Process. Syst.*, Curran Associates Inc., Dec. 2012, pp. 1097–1105.

[15] D. P. Kingma, and J. Ba, "Adam: A Method for Stochastic Optimization," in *International Conference on Learning Representations,* 2015.

[16] S. Chan, "deconvtv - fast algorithm for total variation deconvolution," *MATLAB Central File Exchange*, retrieved October 15, 2020. Available: https://www.mathworks.com/matlabcentral/fileexchange/43600-deconvtv-fast-algorithm-for-total-variation-deconvolution.